\documentclass[openacc]{rsproca_new}

\newtheorem{theorem}{\bf Theorem}[section]

\newtheorem{lemma}{\bf Lemma}[section]

\usepackage{caption}
\usepackage{subcaption}
\usepackage{mathrsfs}
\usepackage{stmaryrd}

\usepackage{numprint}
\npdecimalsign{\ensuremath{.}}
\newcommand{\np}[1]{\numprint{#1}}

\newcommand{\dre}{DRE}		

\newcommand{\spop}[2]{\ensuremath{{#2}_{#1}}}
\providecommand{\SpecSet}{\ensuremath{\mathscr{S}}} 
\providecommand{\ReacSet}{\ensuremath{\mathscr{R}}}

\providecommand{\DecompMax}{\ensuremath{\overline{O}}}
\providecommand{\ratef}{\ensuremath{f}}

\newcommand{\tracked}[1]{\ensuremath{\llbracket #1 \rrbracket}}
\newcommand{\state}{\ensuremath{\sigma}}
\newcommand{\RealSpecies}{\ensuremath{\mathbb{R}^\SpecSet}}
\newcommand{\IntegerSpecies}{\ensuremath{\mathbb{N}^\SpecSet}}
\newcommand{\oi}{\ensuremath{o'}}

\begin{document}

\title{Improved estimations of stochastic chemical kinetics by finite state expansion}

\author{
Tabea Waizmann\,$^{1}$, Luca Bortolussi\,$^{2}$, Andrea Vandin\,$^{3}$, and Mirco Tribastone\,$^{1}$
}

\address{%
$^{1}$IMT School for Advanced Studies, Lucca, 55100, Italy, \\
$^{2}$Department of Mathematics and Geosciences, University of Trieste, 34127, Italy, and \\
$^{3}$Sant'Anna School of Advanced Studies, Pisa, 56127, Italy.}

\subject{systems theory, computational biology}

\keywords{stochastic reaction networks, reaction rate equations, mean approximations, master equation, continuous-time Markov chains}

\corres{Mirco Tribastone\\
\email{mirco.tribastone@imtlucca.it}}

\begin{abstract}
Stochastic reaction networks are a fundamental model to describe interactions between species where random fluctuations are relevant. The master equation provides the evolution of the probability distribution across the discrete state space consisting of vectors of population counts for each species. However, since its exact solution is often elusive, several analytical approximations have been proposed. The deterministic rate equation (DRE) gives a macroscopic approximation as a compact system of differential equations that estimate the average populations for each species, but it may be inaccurate in the case of nonlinear interaction dynamics.
Here we propose finite state expansion (FSE), an analytical method mediating between the microscopic and the macroscopic interpretations of a stochastic reaction network by coupling the master equation dynamics of a chosen subset of the discrete state space with the mean population dynamics of the DRE. An algorithm translates a network into an expanded one where each discrete state is represented as a further distinct species. This translation exactly preserves the stochastic dynamics, but the DRE of the expanded network can be interpreted as a correction to the original one. The effectiveness of FSE is demonstrated in models that challenge state-of-the-art techniques due to intrinsic noise, multi-scale  populations, and multi-stability. 
\end{abstract}


%
%
%
%
%
%
%
%


\maketitle

\section{Introduction}
Stochastic reaction networks are a fundamental model to analyze species that interact stochastically through reaction channels according to dynamics governed by  the well-known master equation~\cite{kampen01}. This provides a microscopic description in terms of a set of coupled linear differential equations,  each defining the time course of a discrete state of the system as a vector of population counts of the species involved. It is widely understood, however, that the master equation is intractable in general, since analytical solutions are available only in special cases and direct numerical integration is hindered by the combinatorial growth of the state space as a function of the abundances of the species. 
Alternatively, it is possible to analyze the network by means of stochastic simulation, e.g.~\cite{Gillespie77}. However, this may preclude other important studies such as stability, perturbation analysis, bifurcation, and parameter inference~\cite{1751-8121-50-9-093001,10.1371/journal.pcbi.1005030}. Thus, it is often useful to consider analytical approximations that trade off precision with computational cost~\cite{MacNamara:2009aa}.

The deterministic rate equation ({\dre}) provides a macroscopic dynamical view by associating one ordinary differential equation with each species representing its mean population. This is exact if each reaction's propensity function is linear, for instance  in monomolecular chemical reaction networks~\cite{Gillespie77}, and for certain classes of networks with bimolecular reactions~\cite{PhysRevE.92.042124}. In general, however, with nonlinear propensities functions, the {\dre} does give the true expectations only in the thermodynamic limit under mild conditions~\cite{kurtz-chem}.
In this case, away from this asymptotic regime the {\dre}s are only an approximation to the true mean dynamics. This occurs, for instance, in models of cell regulation that depend on low-abundance species (in the order of a few units) to describe the behavior of genes~\cite{doi:10.1002/bies.950171112}. Processes such as activation and deactivation that vary with time as a result of various interactions may introduce significant variability in gene expression~\cite{Elowitz16082002}, caused by inherent stochasticity in the bio-molecular processes involved~\cite{PAULSSON2005157,Swain12795}. Since such forms of noise are not accounted for in the \dre, approximation errors may be large.    

Here we present finite state expansion (FSE), an analythical method that keeps track discretely of only a user-defined subset of the state space while collapsing the rest as a continuous approximation. In particular, the state space to be tracked discretely is determined by a parameter that specifies the maximum allowed population level for each species.
FSE is a systematic translation of a stochastic reaction network with arbitrary propensity functions into an expanded one with additional species and modified reactions. Specifically, each tracked discrete state is represented as a new auxiliary species; the original set of reactions is transformed such that the dynamics of the auxiliary species are coupled with those of the original species, whose  role is to buffer the probability mass that falls out of the state space that is tracked.  

FSE enjoys two useful properties. The first concerns its \emph{soundness}, in the sense that any expanded network is stochastically equivalent to the original one:
the state space of the expanded network can be projected onto a lower-dimensional one that still satisfies the Markov property, according to~\cite{BuchholzOrdinaryExact}, and which turns out to correspond to the original network.  Such correspondence does not carry over to the respective DREs of the original and of the expanded networks.  Indeed, any expansion arising from a strict subset of the discrete state space will lead to a DRE with more equations, which can be interpreted as refining terms for the mean estimates. 
Our second theoretical contribution is a result of \emph{asymptotic correctness}, stating that if every discrete state is tracked then the DRE of the expanded network corresponds to the master equation. 

There are several analytical approaches that can be used to improve the accuracy of the \dre. These include moment-closure approximations~\cite{Kuehn2016}, the effective mesoscopic rate equation, which adds correction terms to van Kampen's well-known system size expansion~\cite{doi:10.1063/1.3454685,6392668}, and hybrid techniques~\cite{Hasenauer2014}; \cite{1751-8121-50-9-093001} offers an up-to-date review. However, they are applicable  under certain assumptions such as smoothness of the propensity functions~\cite{doi:10.1063/1.2408422,doi:10.1063/1.3454685,doi:10.1063/1.4802475}, mass-action kinetics~\cite{lee:134107,SOTIROPOULOS2011268,4762250,5605237,Smadbeck14261}, specific network structures, e.g., to describe gene regulatory systems~\cite{Thomas201400049,lma}, and species that can be partitioned into low-abundance and high-abundance classes~\cite{doi:10.1137/110821500,hybrid-stoc-det-cme2011,Hasenauer2014}.
FSE, instead, can be applied to networks with propensity functions of arbitrary form. Using selected case studies from the literature we show how FSE can provide accurate mean estimates in model instances that challenge state-of-the-art methods.

\section{Background theory}
We briefly review here the preliminary definitions and notation on stochastic reaction networks used in the paper.

Consider a set of species $\SpecSet$. Then, $\mathbb{N}^\SpecSet$ and $\mathbb{R}^\SpecSet$ are the sets of all integer and real-valued vectors, respectively, with coordinates represented by the elements in $\SpecSet$. For a given vector $\sigma \in \mathbb{R}^\SpecSet$ (or $\sigma \in \mathbb{N}^\SpecSet$), we denote by $\sigma_S$ the value of the component corresponding to species $S \in \SpecSet$. 
We generalize binary operations to the case where operands $\sigma$ and $\mu$ are such that  $\sigma\in \mathbb{R}^{\SpecSet_1}$ and $\mu\in\mathbb{R}^{\SpecSet_2}$, with $\SpecSet_1 \neq \SpecSet_2$: each binary operation treats them as elements of $\mathbb{R}^{\SpecSet_1\cup\SpecSet_2}$. 

Formally, we denote a reaction network as a pair $(\SpecSet,\ReacSet)$, where $\ReacSet$ is a set of reactions. Each reaction is provided as a triple in the form 
\begin{equation}\label{eq:orig.reaction}
    \rho \xrightarrow{\ratef} \pi,
\end{equation}
where $\rho \in \mathbb{N}^\SpecSet$ are the \emph{reactants}, $\pi  \in \mathbb{N}^\SpecSet$ are the \emph{products}, and $\ratef$ is the propensity function, $\ratef: \mathbb{R}^\SpecSet \to \mathbb{R}_0^+$, with arbitrary form. We will use the standard notation for reactants and products whereby only the nonzero components are written out, separated by the plus sign. For instance, given the species $\SpecSet = \{ A,B,C \}$, Eq.~\ref{eq:orig.reaction} corresponds to the reaction 
\begin{equation}\label{eq:sample}
A + 2B \xrightarrow{f} C
\end{equation}
when $\rho_A = 1$, $\rho_B = 2$, $\rho_C = 0$ and $\pi_A = \pi_B = 0$, $\pi_C = 1$. 
A discrete state of a reaction network is described by a vector $\sigma \in \mathbb{N}^\SpecSet$, where $\sigma_S$ denotes the population of species $S$ in that state. Then, $f(\sigma)$ is the parameter of the exponential distribution of the firing time of that reaction. Upon firing, the system may transition from state $\state$ to $\state + \pi - \rho$, thus defining a stochastic behavior in terms of a Markov jump process. Its dynamics is defined by the master equation. It gives the probability $P_{\state}(t)$ of finding the Markov chain in state $\state$ at time $t$: 
\[
\frac{dP_{\state}(t)}{dt} = \sum_{\rho \xrightarrow{~f~} \pi} -f(\state) P_{\state}(t) + f(\state + \rho - \pi) P_{\state + \rho - \pi}(t) .
\]  

Formally, the master equation may be defined for all $\sigma \in \IntegerSpecies$. However, its solution will be nonzero only for those states that are reachable from the states that have nonzero probability at time $t = 0$. The reachable set of states, also called the \emph{state space}, can be defined as the smallest set such that the following hold:
\begin{enumerate}
    \item $\sigma$ is in the reachable set if $P_\sigma(0) > 0$;
    \item $\sigma$ is in the reachable set if $\sigma'$ is in the reachable set and there exists a reaction $\rho \xrightarrow{f} \pi$ such that $\sigma' + \pi - \rho = \sigma$.
\end{enumerate}

For simplicity (and without loss of generality) we will consider networks where the initial probability distribution $P_\sigma(0)$ is concentrated in one state only, which is called the \emph{initial state}.  Additionally we shall restrict to \emph{well-defined}  reaction networks where each propensity function evaluates to zero for all multisets that do not have the minimum population counts described by the reactants. Formally, a reaction network is well-defined if every reaction $\rho \xrightarrow{\ratef} \pi$ is such that  $f(\sigma)=0$ if $\rho > \sigma$ (the inequalities shall be intended component-wise from now on).
This guarantees that the Markov chain does not reach states with negative population counts.

The state space, hence the number of equations required for stochastic analysis, may be finite or infinite depending on the network stoichiometries. Even in the case of finite state spaces, its size may grow combinatorially large with the population counts of the initial state. This may practically preclude exact analysis in most models of interest. The \dre\ provides a compact model with $| \SpecSet |$ variables. Each variable approximates the expected population level of each species at time $t$, denoted by the vector $X(t) \in \RealSpecies$, as the solution of the system:
\[
\frac{dX(t)}{dt} = \sum_{\rho \xrightarrow{~f~} \pi} f(X(t)) (\pi - \rho) .
\]
The true expected population counts, denoted by $\mathbb{E}[Y]$, are known to satisfy
\[
\frac{d\mathbb{E}[Y(t)]}{dt} = \sum_{\rho \xrightarrow{~f~} \pi} \mathbb{E}[f(Y(t))] (\pi - \rho). 
\]
However this system is not self-consistent because there are no equations for the expected values of the propensity functions appearing in the right-hand sides. 
The DRE \emph{closes} the true equations for the expected values by replacing $\mathbb{E}[f(Y(t))]$ with $f(\mathbb{E}[Y(t)])$, introducing an approximation error if the propensity functions are not linear. Under mild conditions such error is known to vanish asymptotically when the initial population levels go to infinity and the DRE is understood as a system of re-scaled equations for the concentrations of species, rather than absolute population counts~\cite{kurtz-chem}. 

\section{Methodology}

\subsection{Finite state expansion}

With FSE, the original set of species $\SpecSet$ of a reaction network is meant to represent the continuous dynamics. This is expanded with a set of auxiliary species, each tracking a specific discrete state.
The auxiliary species are defined by the user through an upper bound to the population count to be tracked discretely for each species. Thus, in effect FSE yields a lattice of expansions depending on the choice of the upper bounds. 

Let us denote by  $\DecompMax \in \mathbb{N}^\SpecSet$ such upper bound, where each component $\DecompMax_S$ gives the maximum abundance to be tracked for species $S$. For each discrete state $o \leq \DecompMax$, we denote by $\llbracket o \rrbracket$ the corresponding auxiliary species that is considered in the expansion.  Thus we may define  $\SpecSet_{\DecompMax}$ to be the set of species in the expanded network as
$\SpecSet_{\DecompMax} = \SpecSet \cup \left\{ \llbracket o \rrbracket \mid o \leq \DecompMax \right\}$.
For example, in a network with the single reaction as in Eq.~\ref{eq:sample}, let us choose $O_A = O_B = O_C = 1$. Then, the expanded network will have auxiliary species 
$\llbracket A \rrbracket$, $\llbracket B \rrbracket$, $\llbracket C \rrbracket$, $\llbracket A + B \rrbracket$, $\llbracket A + C \rrbracket$, $\llbracket B + C  \rrbracket$, $\llbracket A+B+C \rrbracket$, and $\llbracket\mathbf{0} \rrbracket$, where the last species denotes the zero vector being tracked. 
We remark that, similarly to the definition of the master equation, it is convenient to consider all states within the upper bound when describing the theory. However, also in this case, not all discrete states may be reached depending on the stoichiometry of the reaction network, hence they can be removed in practice during the analysis.

FSE replaces each reaction $\rho \xrightarrow{f} \pi$ in the original network with a set of reactions for each tracked state $\llbracket o\rrbracket$ as follows:  
\begin{equation}\label{eq:expansion}
{\llbracket o\rrbracket} + \eta \xrightarrow{~f_o~} {\llbracket o'\rrbracket} + \psi , \qquad \text{for~} o \leq \DecompMax, 
\end{equation}
where $\eta$ and $\psi$ are vectors of original species, ${\llbracket o'\rrbracket}$ is a target auxiliary species, and $f_o$ is a modified propensity function that accounts for the fact that the reactant auxiliary species is ${\llbracket o'\rrbracket}$. Specifically, we have:
\begin{equation}\label{eq:eta}
\begin{split}
\eta & = \max(\mathbf{0}, \rho - o) \\ 
\psi & = \max(\mathbf{0}, \max\left( \mathbf{0}, o - \rho \right) + \pi -  \DecompMax) \\
o' & = \min(\DecompMax, \max\left( \mathbf{0}, o - \rho \right) + \pi).  
\end{split}
\end{equation}
Intuitively, for each original reaction as in Eq.~\ref{eq:orig.reaction}, Eq.~\ref{eq:expansion} conditions its dynamics with respect to $\llbracket o \rrbracket$ being the discrete state being tracked. Any expanded reaction maintains the same overall counts of reactants and products as the originating reaction, with a product tracked state $\llbracket o' \rrbracket$ that results from the addition of products and removal of reactants within the upper bound $\DecompMax$; $\eta$ and $\psi$ act as \emph{buffer species} for populations that are not explicitly tracked. Finally, the propensity function $f_o$ is derived from the original one $f$  as 
\begin{equation}\label{eq:exp.fun}
f_o:  \mathbb{R}^{\SpecSet_{\DecompMax}} \rightarrow \mathbb{R}_0^+,  \quad \text{with} \quad  
 f_o( x) = \spop{{\llbracket o\rrbracket}}{x} f(o+ x_{|\SpecSet}).
\end{equation} 
This modification accounts for the fact that the tracked species $\llbracket o \rrbracket$ encodes additional population counts, as given by $o$. 
 
For example, let us consider an expansion for the reaction in Eq.~\ref{eq:sample} assuming that it evolves with mass-action kinetics. In general, for a reaction with reagents $\rho$ and kinetic parameter $k > 0$, the propensity function by mass-action kinetics for state $\sigma$ is given by $f_k(\sigma) =  k \prod_{S}\binom{\state_S}{\rho_S}$. Here the propensity function is
\[
f(x) = k x_A x_B (x_B - 1)/2.
\]
Assuming that the upper bounds are $\DecompMax_A = \DecompMax_B = \DecompMax_C = 1$, the expansion for the tracked state $\llbracket A + B \rrbracket$ is given by 
 \[
 \llbracket A + B \rrbracket + B \xrightarrow{f'} \llbracket C \rrbracket. 
 \]
 Since the tracked state $\llbracket A + B \rrbracket$ does not have enough copies of $B$, one further copy is used by the buffer species. The product of this reaction does not involve any of the buffer species because $\llbracket C \rrbracket$ is within the chosen bounds. By Eq.~\ref{eq:exp.fun}, $f'$in the expanded reaction is
 \[
 \begin{split}
 f'(x) & = k x_{\llbracket A + B \rrbracket} (1 + x_A) (1 + x_B) \big((1 + x_B) -1 \big)/2 \\
 & = k x_{\llbracket A + B \rrbracket} (1 + x_A) (1 + x_B) x_B /2.
 \end{split}
 \]
We denote by $\ReacSet_{\DecompMax}$ the full set of reactions in the expanded network. 

Every expansion is stochastically equivalent to the original network, in the sense that there is a unique marginal probability distribution for the overall population of each species at every time point. 
This equivalence can be stated  in the sense of ordinary lumpability for Markov chains~\cite{BuchholzOrdinaryExact}. That is, with the master equation we show that the probability of being in a state in the original reaction network equals the sum of the probabilities across all states in the expanded network with the same overall abundances for each species. This relation holds at all time points, provided that it  is satisfied for the respective probability distributions at time 0.\footnote{The proofs of the forthcoming theorems are available in the enclosed Appendix, Section~\ref{sec:proofs}. 
} 
\begin{theorem}\label{thm:stochEqui}
Let $P$ and $\hat{P}$ denote the solutions of the master equation in the original and expanded network, respectively. Then, for all $t$, 
\[
\sum_{o + \xi = \sigma} \spop{\llbracket o\rrbracket  + \xi}{\hat{P}}(0) = \spop{\sigma}{P}(0)
 \implies  \sum_{o + \xi = \sigma} \spop{\llbracket o\rrbracket  + \xi}{\hat{P}}(t) = \spop{\sigma}{P}(t).
\]
\end{theorem}

By construction, if $\DecompMax = \mathbf{0}$ then the original and expanded networks coincide. The other limit case, when the auxiliary set of species  contains all discrete states, corresponds to a fully expanded reaction network. In this case, the DRE of the expanded network corresponds to the master equation of the original network, hence no approximation occurs. 



\begin{theorem}\label{thm:dre}
Consider a well-defined reaction network $(\SpecSet,\ReacSet)$ and let $(\SpecSet_{\DecompMax}, \ReacSet_{\DecompMax})$ be its expansion. 
Let $X(t)$ be the \dre\ solution of the expanded network and $P(t)$ the solution of the master equation of the original network at time $t$.  
Then, the following hold:
\begin{enumerate}
\item if $X_S(0)= 0$ then $X_S(t)= 0$ for all $t$ and $S \in \SpecSet$;
\item if $X_{\tracked{o}}(0) = P_o(0)$ then $X_{\llbracket o \rrbracket}(t) = P_o(t)$, for all $t$ and $o \in \IntegerSpecies$.
\end{enumerate}
\end{theorem}


\begin{figure*}
\begin{center}
\includegraphics[width=1.0\textwidth]{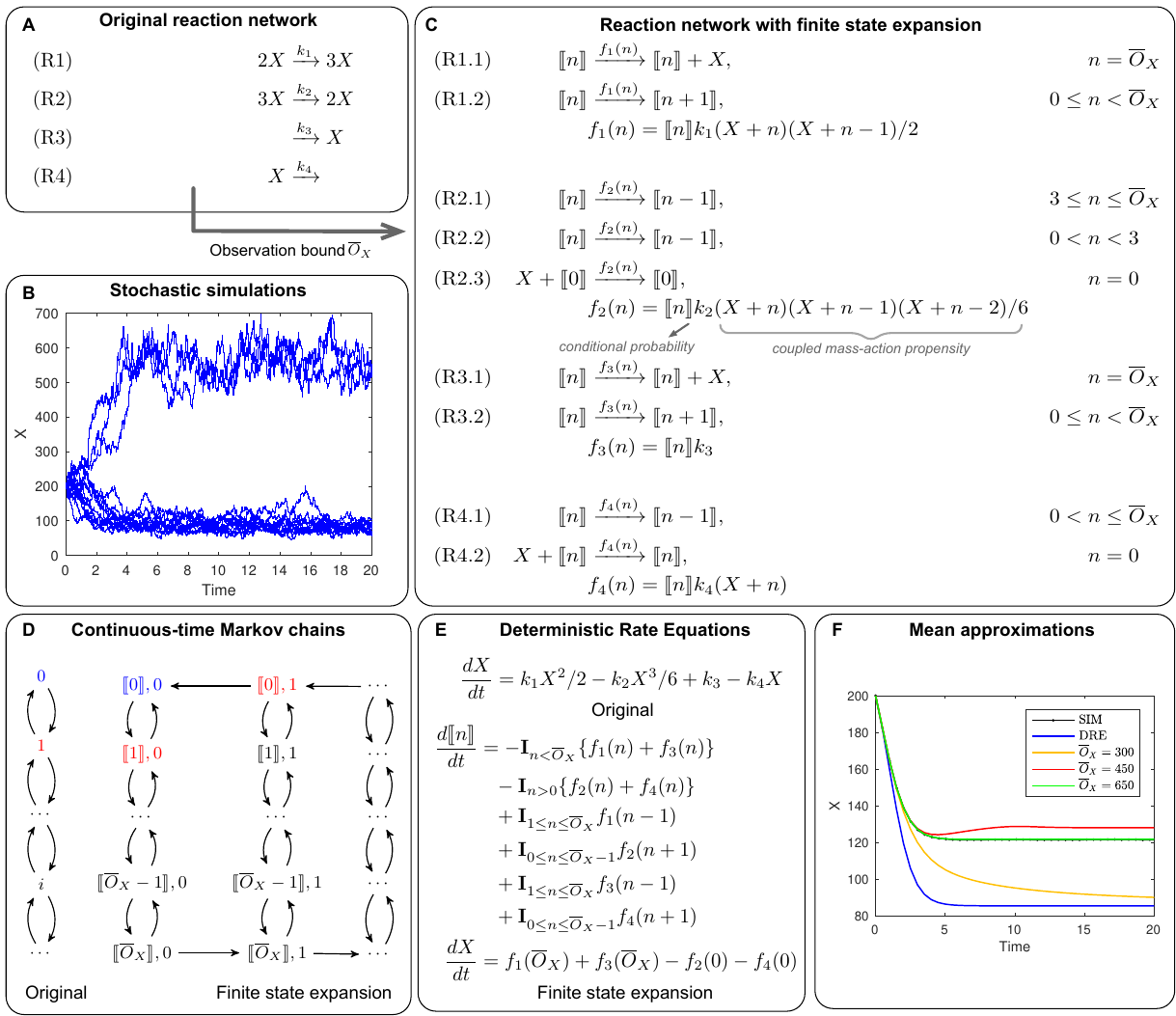}
\end{center}
\caption{{\bf The FSE method applied to the Schl\"ogl system~\cite{Schloegl1972}.} (A) Mass-action reactions with kinetic parameters taken from ref.~\cite{doi:10.1021/bp070255h}: $k_1 = 0.03$, $k_2 = 0.0001$, $k_3 = 200$, $k_4 = 3.5$. (B)~Stochastic simulations show the well-known bimodality of the steady-state probability distribution of species $X$. (C)~For a given upper bound $\DecompMax_X$ on the population of species $X$ to be tracked explicitly, FSE yields the auxiliary species denoted by $\tracked{0}$, $\tracked{1}$, \ldots, $\tracked{\DecompMax_X}$. The original species $X$ acts as buffer that collects untracked populations levels. For example, reaction R1.1 derives from reaction R1 when the autocatalytic formation of a new molecule occurs when the system tracks the discrete state $\tracked{\DecompMax_X}$, thus requiring to increase the buffer species $X$ by one element. 
Even when the system tracks a discrete state that does not require buffering (R1.2), the propensity function $f_1(n)$ of the reaction effectively considers an overall kinetics of mass-action type, since the factor $k_1(X+n)(X+n-1)/2$ models the total rate due to number of possible collisions between pairs of $X + n$ indistinguishable molecules. Intuitively, the factor $\tracked{n}$ conditions these events to the system tracking $n$ discrete molecules. (D)~The original state space counts the number of copies of $X$. The state space in the expanded network consists of the pair tracked discrete state/population level of the buffer species. By Theorem~\ref{thm:stochEqui}, the sum of the probabilities across all pairs that have the same overall population matches the corresponding probability in the original Markov chain (as exemplified by matching colors of the states). (E)~The single-dimensional DRE of the original Schl\"ogl model is expanded into a DRE with $\DecompMax_X + 1$ variables (where $\mathbf{I}$ denotes the indicator function); an estimate of the total mean population at time $t$ can be computed as $X(t) + \sum_n n \cdot \tracked{n}(t)$. F) Starting from a population of 200 elements of $X$, the original bi-stable DRE converges to one equilibrium at 85.50 (blue line). {FSE} achieves excellent agreement with an upper bound $\DecompMax_X = 650$ (with respect to the average computed by stochastic simulation with \np{100000} runs).}\label{fig:fse.example}
\end{figure*}

\section{Results}

With a number of case studies from the literature, here we show that FSE can refine the accuracy of mean estimates of species populations, even with modest expansions. Ground-truth mean trajectories were computed by stochastic simulation via Gillespie's algorithm\cite{Gillespie77}. The numerical experiments herein reported were performed with an implementation of {FSE} within the tool ERODE\cite{tacas17}, publicly available  at \href{https://www.erode.eu}{https://www.erode.eu}.  

\subsection{Schl\"ogl system}

The well-known Schl\"ogl system is an autocatalytic process for a single species $X$\cite{Schloegl1972}. The DRE of the original Schl\"ogl model has two equilibrium points, owing to its strong (cubic) nonlinearity\cite{BISHOP20101}, deterministically converging only to one\cite{Vellela925}. Its discrepancy with respect to the average mean trajectory computed by stochastic simulation has been observed for a long time\cite{doi:10.1063/1.459735}. 
Fig.~\ref{fig:fse.example} provides a fully worked application of our FSE as a function of the upper bound for $X$. The solutions to the DRE of the expanded networks show that larger values of such upper bound increasingly improve the accuracy of the mean estimates.

\subsection{Heterodimerization model}\label{sec:hetero}


\begin{figure}
\hspace{-0.5cm}
	\includegraphics[scale=0.8]{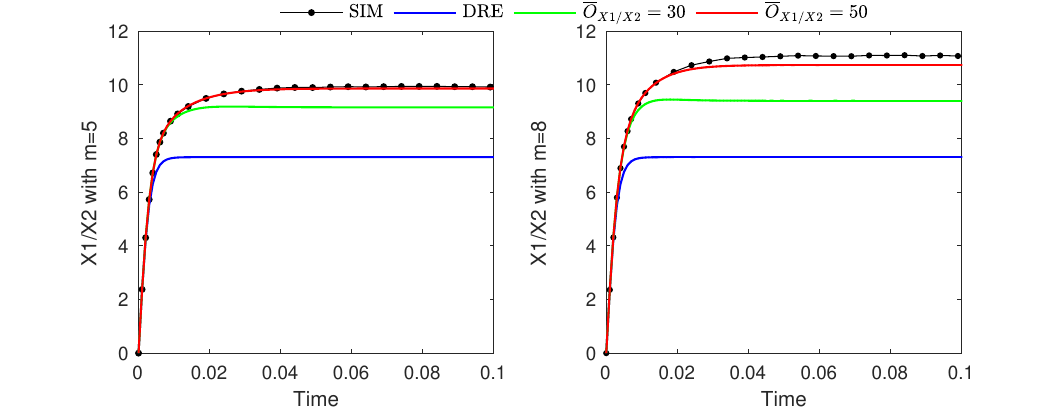}
\caption{{\bf  Heterodimerization model (Eq.~\ref{eq:heterodim}).} Numerical simulations comparing stochastic simulation (\np{300000} runs), \dre\ and FSE for different upper bounds $\DecompMax_{X_1/X_2}$ at burst sizes $m=5$ (left) and $m=8$ (right). The DRE approximation is unaffected by the variation of $m$ while the true population averages are increased at higher burst sizes. Corrections for species $X_3$ are similar with a generally smaller error (cf. Fig.\ref{fig:sen:hetero}).
Kinetic parameters were set as follows: $k_1 = 2500/m$, $k_2 = 40$, $k_3 = 50$. For a burst size of $m = 5$, the parameters are as in Ref.~\cite{doi:10.1063/1.3454685}. Initial condition was the zero state.
}\label{fig:heterodim_2in1}
\end{figure}
We now consider a model from Ref.~\cite{doi:10.1063/1.3454685}, where the main source of noise is the variance caused by the production of the two species $X_1$ and $X_2$ undergoing heterodimerization occurs in bursts of size $m$:
\begin{align}
 & \xrightarrow{k_1} mX_1 & & \xrightarrow{k_1} mX_2 & X_1 + X_2 &  \xrightarrow{k_2} X_3 \label{eq:heterodim} \\ 
  X_1 & \xrightarrow{k_3} & X_2 & \xrightarrow{k_3} & X_3 & \xrightarrow{k_3} \notag 
\end{align} 
We study two cases with $m=5$ and $m=8$; for a better comparison the influx rates were kept constant by setting $k_1 = 2500/m$. Because of the symmetries between $X_1$ and $X_2$ we consider equal observation bounds for them. The observation bound on $X_3$ was set to zero. In this model, the DRE is insensitive to the choice of $m$ while the stochastic trajectories do depend on the burst size. Since a larger $m$ introduces more noise, larger observation bounds are needed to increase the accuracy of the approximation (Fig.~\ref{fig:heterodim_2in1}). 

\subsection{Protein degradation model}\label{sec:degradation}

\begin{figure}
\hspace{-0.6cm}
	\includegraphics[scale=0.8]{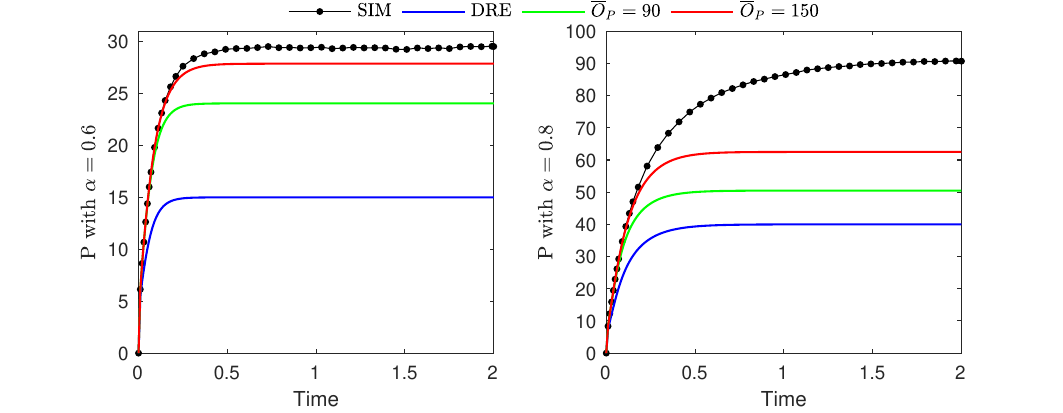}
\caption{{\bf Protein degradation model (Eq.~\ref{eq:grima_C}).} Numerical simulations comparing stochastic simulation (\np{200000} runs), \dre\ and FSE for different upper bounds $\DecompMax_{P}$ at enzyme saturation levels $\alpha=0.6$ (left) and $\alpha=0.8$ (right). Enzyme observation bounds have been fixed at $\DecompMax_E=\DecompMax_C=60$ to encompass the total population $E_T$.
Kinetic parameters were set as follows: $k_2 = 4$, $k_3 = 3$, $k_4 = 37$, $k_5 = 10$, $k_1 = \alpha k_4 E_T/m = 74\alpha$. The initial state is $(P,E,C,Pd)= (0,60,0,0)$. 
}\label{fig:grima_C_2in1}
\end{figure}
The model of enzyme-catalysed protein degradation from~\cite{doi:10.1063/1.3454685} allows us to study the behavior of FSE when more species are to be tracked with nonzero population bounds than in the previous two case studies. 
Here, protein $P$ is generated in bursts of size $m$ and can bind to catalyst enzyme $E$, forming the enzyme-substrate complex $C$. When the protein unbinds from the enzyme, it can be degraded, forming $P_d$. The total amount $E_T$ of catalyst enzyme in the system always remains constant:
\begin{align}
 & \xrightarrow{k_1} m P & & & P_d & \xrightarrow{k_5} \notag \\
 P+E & \xrightarrow{k_2} C  & C & \xrightarrow{k_3} P+E & C & \xrightarrow{k_4} P_d+E \label{eq:grima_C} 
\end{align} 
By varying the burst production rate $k_1$, different saturation levels of the catalyst enzyme are reached. The closer the ratio $\alpha= mk_1/k_4E_T$ is to $1$, the more saturated the enzyme becomes with substrate.

The parameters given in \cite{doi:10.1063/1.3454685} for this model assume a burst size of $m=30$ and a total enzyme population of $E_T=60$. For an accurate approximation, FSE requires nonzero observation bounds for $E$, $C$, and $P$ (Fig.~\ref{fig:grima_C_2in1}); similarly to the previous case, no observation bound is required for species $P_d$. The size of the tracked state space grows proportionally to the product  $\DecompMax_E \cdot \DecompMax_C \cdot \DecompMax_P$. 
Additionally, in contrast to the previous model, burst production rates are not adjusted when increasing the burst size here. The overall higher production rate increases the population additionally to the effect caused by higher variances. This exacerbates the need for higher observation bounds at higher burst sizes.  For a fixed choice of the observation bounds, larger values of $\alpha$ tend to worsen the accuracy of the FSE approximation.

\subsection{Feedback switch} 

\begin{figure}
\hspace{-0.6cm}
\includegraphics[scale=0.8]{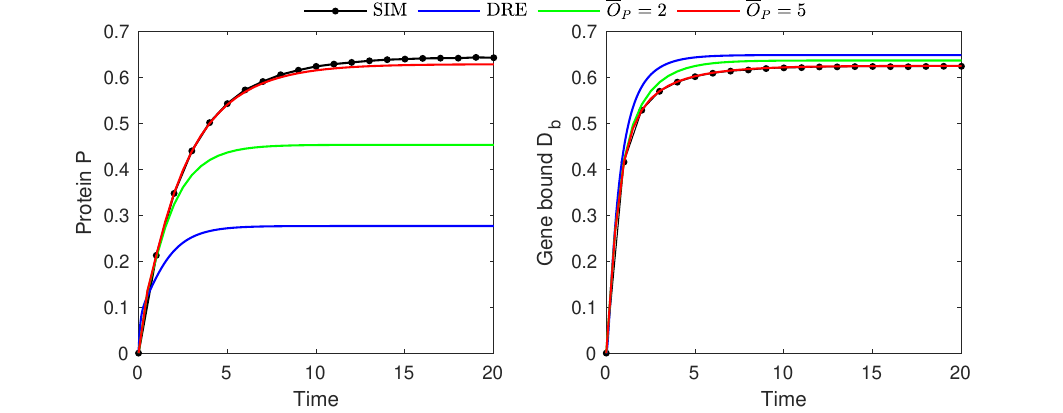}
\caption{{\bf  Genetic feedback switch in Eq.~\ref{eq:feedback_switch}.} Numerical simulations comparing stochastic simulation (\np{1000000} runs), \dre\ and FSE for fixed $\DecompMax_{D_u} = \DecompMax_{D_b}= 1$ and different upper bounds $\DecompMax_{P}$. The resulting \dre\ from {FSE} has $2\cdot\DecompMax_{P} + 2$ equations. 
Kinetic parameters were set as follows: $r_u = 1.0$, $r_b = 0.5$, $k_f = 0.1$, $k_b = 1.0$, $s_b = 10.0$, $s_u = 0.5$. The initial state is $(P, D_u, D_b) = (0,1,0)$.}\label{fig:feedback_switch}
\end{figure}

Let us now consider a model of a genetic feedback switch taken from~\cite{PhysRevE.72.051907} and~\cite{doi:10.1063/1.4736721}:
\begin{align}
D_u & \xrightarrow{r_u} D_u + P &
D_b & \xrightarrow{s_u} D_u + P \notag \\
D_b & \xrightarrow{r_b} D_b + P &  
D_u + P & \xrightarrow{s_b} D_b \label{eq:feedback_switch} \\
D_b & \xrightarrow{k_b} D_u &
 P & \xrightarrow{k_f} \notag
\end{align} 
Species $D_u$ and $D_b$ represent the state of a single gene when its promoter region is unbound (respectively, bound) to a protein $P$. The reaction propensities obey mass-action dynamics through the kinetic parameters on the arrows.
This is a basic model for negative autoregulation, a well-known motif appearing in more than 40\% of the known transcription factors in \emph{E.coli}\cite{Shen-Orr:2002aa}. A natural choice of upper bounds for the gene species is $\DecompMax_{D_u} = \DecompMax_{D_b} = 1$, by which the \dre\ of the expanded network can be interpreted as the solution of the conditional expectation of the protein population based on the gene state. Small values of $\DecompMax_{P}$ yield a significant correction of the protein levels as well as of the marginal probability distribution of the gene state (Fig.~\ref{fig:feedback_switch}). 

\subsection{Toggle switch} 
The toggle switch network is a fundamental regulatory system of two mutually repressing genes\cite{minetwork}. Its mathematical modeling is challenging because of multimodality\cite{Tian8372,Thomas201400049}, as well as stochastic noise due to the species such as mRNA present in low molecular abundances\cite{Kaern:2005aa}. Here we study the reaction scheme analyzed in\cite{doi:10.1063/1.4905196}, consisting of a mass-action variant from~\cite{minetwork}:
\begin{align}
  \!\! & \xrightarrow{k_1}  M_i &  \!\!\!\! M_i & \xrightarrow{k_2}  & \!\!\!\! M_i & \xrightarrow{k_3} S_i  \notag \\ 
 \! \!\!\!\! S_i & \xrightarrow{k_4} S_i + P_i & \!\!\!\! S_i & \xrightarrow{k_5}  &   \!\!\!\! P_i & \xrightarrow{k_6} \emptyset, \label{eq:ts} \\ 
 \! \!\!\!\!  & & \!\!\!\! S_i + M_j & \xrightarrow{k_7} S_i, &  & \!\! i,j \in \{A,B\}, \ i \neq j, \notag 
\end{align}    
where $M_i$ and $S_i$ denote the precursor mRNA and the mRNA for target protein $P_i$. The last two reactions model mutual inhibition by means of a precursor of one protein repressing the mRNA of the other. 

\begin{figure}
\hspace{-0.6cm}
\includegraphics[scale=0.8]{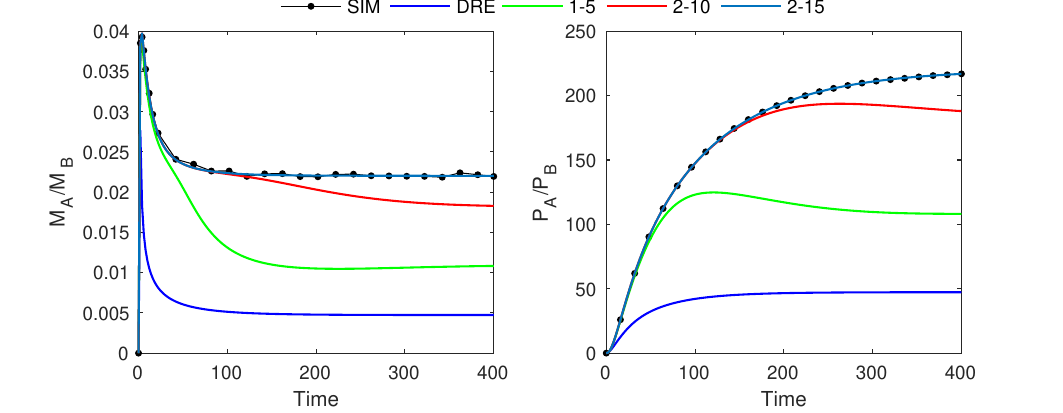}
\caption{{\bf Toggle switch (Eq.~\ref{eq:ts})}. Numerical simulations comparing stochastic simulation (500000 runs), \dre\ and {FSE} by fixing $\DecompMax_{P_A} = \DecompMax_{P_B}= 0$  while using different upper bounds $\DecompMax_M$ and $\DecompMax_S$ ($\DecompMax_M$--$\DecompMax_S$ in short) for the number of copies of $M_A$/$M_B$ and $S_A$/$S_B$ (as indicated in the legend), respectively. Initial condition was the zero state. The size of the DRE for the tested choices of upper bounds is equal to $(\DecompMax_M+1)^2\cdot(\DecompMax_S+1)^2 + 6$ (corresponding to 150, 1095 and 2310 equations for $\DecompMax_M\text{--}\DecompMax_S = 1\text{--}5$, $\DecompMax_M\text{--}\DecompMax_S = 2\text{--}10$, and $\DecompMax_M\text{--}\DecompMax_S = 2\text{--}15$, respectively). Kinetic parameters were chosen as follows: $k_1 = 0.05$, $k_2 = 0.1$, $k_3 =1.0$, $k_4 = 10.0$, $k_5 = 0.01$, $k_6 = 0.1$, $k_7 = 20.0$. Protein production (right plot) is controlled by a low population of precursor mRNA (left plot), which causes significant underestimation errors with \dre. Increasing the upper bounds of {FSE} improves the accuracy of the mean estimate. Corrections for species $S_A$ and $S_B$, not reported here, are similar.}\label{fig:ts}
\end{figure}

When protein production is controlled by low populations of precursor mRNA, the stochastic fluctuations are not adequately approximated by the \dre. By explicitly observing few copies of mRNA (up to tens), FSE provides precise estimates of the time courses of the mean populations (Fig.~\ref{fig:ts}). The resulting systems of equations, of size at most \np{2310}, can be analyzed effectively, as opposed to time-consuming simulations using hybrid stochastic-deterministic approaches such as in\cite{doi:10.1063/1.4905196}.        

\section{Comparison with related work}

Using the models presented in the previous section, here we compare FSE with state-of-the-art analytical techniques  that can be used to obtain approximate estimates of mean population levels in stochastic  reaction networks.  Specifically, we considered the following methods:
\begin{itemize}
    \item Moment-closure approximation (MCA). We considered the second-order low-dispersion moment closure\cite{10.1371/journal.pone.0146732,doi:10.1063/1.4934990}, in which variance and covariance are the highest observed moments and all higher-order central moments are set to zero; in all models considered in this paper, computing approximations with higher-order moments did not improve the quality of the approximation.  
    \item The effective mesoscopic rate equation (EMRE), which adds mean-correction terms to the linear-noise approximation under the assumption of an underlying Gaussian process\cite{doi:10.1063/1.3454685}.
    \item The method of conditional moments (MCM), a hybrid analytical technique combining a discrete representation of low-abundance species and a moment-based approximation of high-abundance ones\cite{Hasenauer2014}. 
    \item Finite state projection (FSP), a well-known method to obtain a finite-dimensional master equation through a truncation of the state space by redirecting transitions toward unobserved states into an absorbing state with provable bounds\cite{doi:10.1063/1.2145882}.
\end{itemize}
For this study, we used an implementation of these techniques as available on the software tool CERENA\cite{10.1371/journal.pone.0146732}.

\begin{figure}
\centering
\includegraphics[scale=0.65]{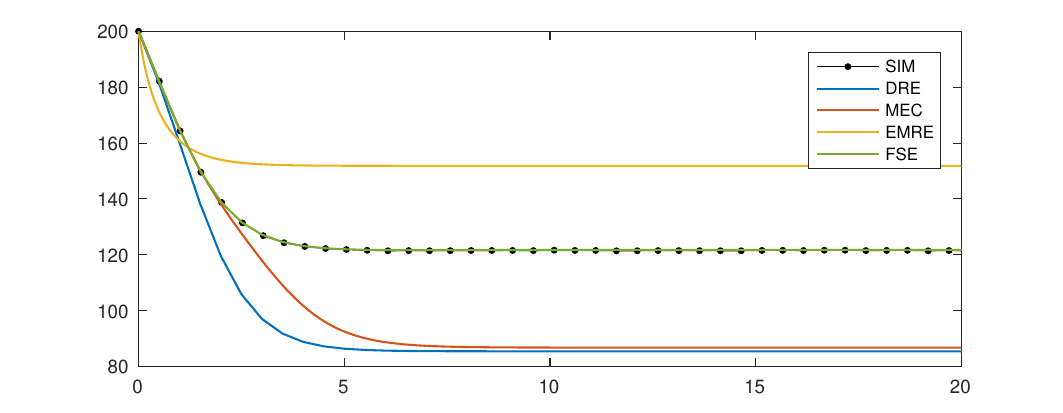}
\caption{{\bf Comparison with related techniques on the Schl\"ogl model.} The average population of $X$ computed by stochastic simulation (100000 runs) is compared against DRE, MCA, EMRE and FSE with $\DecompMax_{X} = 650$.}\label{fig:schloegl_sota}
\end{figure}

\begin{figure}
\centering 
\includegraphics[scale=0.65]{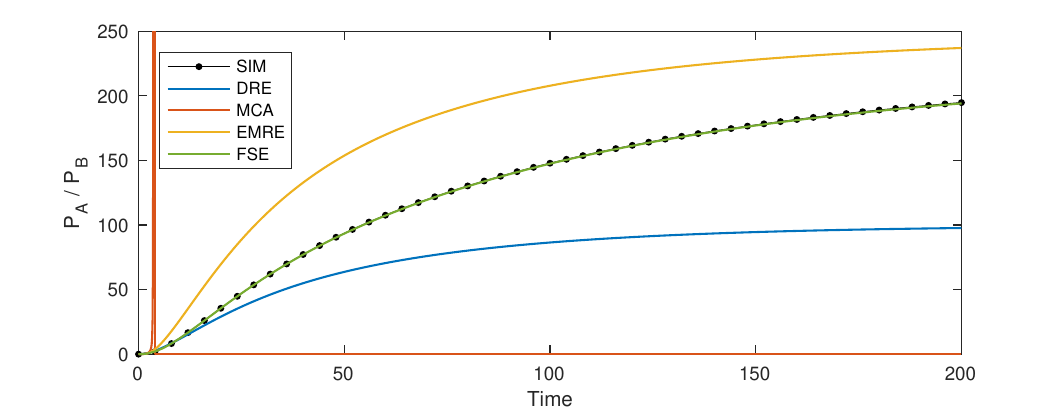}
\caption{{\bf Comparison with related techniques on the toggle switch model.} Stochastic simulation to compute the average populations of species $P_A/P_B$ (\np{500000} runs) is compared against DRE, MCA, EMRE, and FSE. FSE is run with upper bounds $\DecompMax_P = 0$, $\DecompMax_M = 2$ and $\DecompMax_S = 10$. MCA estimates population levels approaching \np{75000} (out of scale in this plot to improve readability) before dropping to zero.}\label{fig:toggle_sota}
\end{figure}

The Schl\"ogl system is known to stress MCA because of their reported difficulties with multimodal distributions\cite{doi:10.1063/1.4929837,doi:10.1063/1.4892838}. Fig.~\ref{fig:schloegl_sota} shows that MCA behaves similarly to DRE in this case, while EMRE tends to overestimate the mean population of species $X$ at longer time horizons. Similar results were obtained on the toggle switch network (Fig.~\ref{fig:toggle_sota}). Here we confirm physically meaningless moment-closure estimates due to the presence of low-abundance species, as already reported in\cite{5605237}. Furthermore,  MCM could not be tested on this model since its implementation returned with an error.  

\begin{figure}
\centering
\includegraphics[scale=0.65]{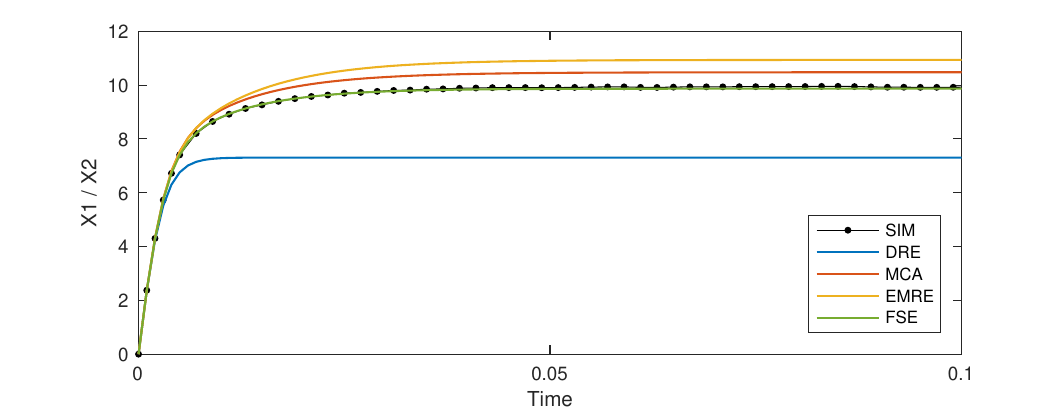}
\caption{{\bf Comparison with related techniques on the heterodimerization model with burst size $m=5$.} Stochastic simulation to compute the average populations of species $X_1/X_2$ (\np{300000} runs) is compared against DRE, MCA, EMRE, FSE using upper bounds $\DecompMax_{X1} =\DecompMax_{X_2} = 50$ and $\DecompMax_{X_3} = 0$. 
}\label{fig:heterodim_sota}
\end{figure}
\begin{figure}
\centering
\includegraphics[scale=0.65]{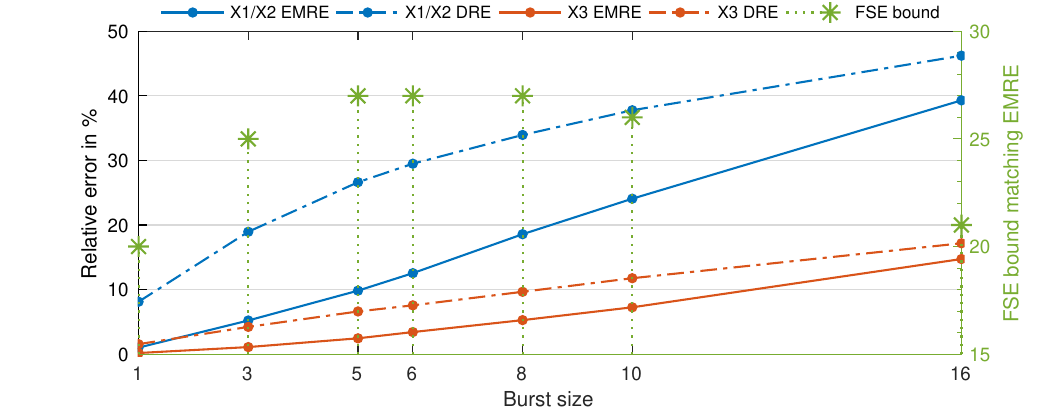}
\caption{{\bf Direct comparison against EMRE on the heterodimerization model across different burst sizes.} 
Although it is consistently better than the DRE (broken line, shown for reference) for all tested values, the relative error of EMRE to the stochastic simulation rises on all species with increasing burst size.
In green, the graph shows for each burst size the FSE bound $\DecompMax_{X_1/X_2}$ that is necessary to match the accuracy of EMRE.
}\label{fig:heterodim_EMRE}
\end{figure}

Both MCA and EMRE provide accurate estimates in the heterodimerization model from Eq.~\ref{eq:heterodim} (Fig.~\ref{fig:heterodim_sota}). 
This model also allows examining the effects of burst size variation without also changing the total production rate. We used this model for a closer comparison to EMRE in this regard.
Figure~\ref{fig:heterodim_EMRE} shows that, similarly to FSE (with fixed observation bounds) and the DRE approximation, the approximation error of EMRE increases with the burst size $m$. For each tested burst size, Figure~\ref{fig:heterodim_EMRE} also marks (secondary $y$-axis) the minimum FSE observation bound $\DecompMax_{X_1/X_2}$ for which the accuracy of EMRE can be matched. In this study, the highest observation bound $\DecompMax_{X_1/X_2}=27$ is enough to match EMRE's accuracy in the range $5\leq m\leq 8$. For burst sizes larger than 8, the loss of accuracy in EMRE outpaces the analogous effect in FSE, to the degree that the bound to match EMRE becomes small. The accuracy of FSE can be increased further by raising the bound above the marked values (Fig.~\ref{fig:sen:hetero}).


The main challenge of the protein degradation model in Eq.~\ref{eq:grima_C} for FSE is the rapid growth of the state space as a function of the observation bounds. Using the observation bounds as in Fig.~\ref{fig:grima_C_2in1}, FSE is outperformed by both EMRE and MCA for $\alpha = 0.8$. For $\alpha = 0.6$, FSE is more accurate than EMRE; however, it uses a significantly more complex system with 9214 differential equations.

\begin{figure}
\centering
\includegraphics[scale=0.7]{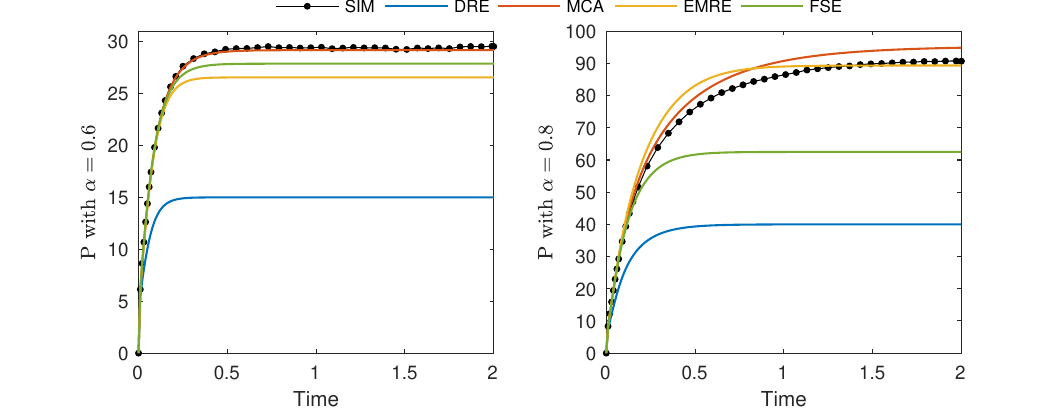}
\caption{{\bf Comparison with related techniques on the protein degradation model.} Stochastic simulation to compute the average populations of species $P$ (\np{200000} runs) is compared against DRE, MCA, EMRE and FSE for two different catalyst saturation levels $\alpha$. FSE is run with upper bounds $\DecompMax_{P} = 150$, $\DecompMax_{E} =\DecompMax_{C} = 60$ and $\DecompMax_{P_d} = 0$.}\label{fig:grima_C_sota}
\end{figure}


In the genetic feedback switch model, species $D_u$ and $D_b$ describe the distinct binary states of a single gene. Hence they represent the natural candidates of the low-abundance class when applying MCM. On this model, however, the method could not return valid results as early as time point $0.36$. We further tested a gene regulatory model with an inhibition feedback loop taken from~\cite{doi:10.1063/1.4986560}, where it was studied using a hybrid stochastic/deterministic method based on piecewise deterministic Markov processes. Here, MCM showed similar difficulties that confirm already reported numerical issues~\cite{1751-8121-50-9-093001}. (The numerical results of this analysis, not shown here, are replicable using the supporting data of this article.) Stochastic models of gene expression such as these are amenable to domain-specific techniques (see, e.g.,~\cite{lma} and references therein), which may prove more effective. For example, in the enclosed Appendix~\ref{sec:lma} we compare FSE against the linear mapping approximation method from~\cite{lma}, showing that FSE requires larger systems of differential equations to obtain a similar level of accuracy.    



\begin{figure}
\centering
\includegraphics[scale=0.65]{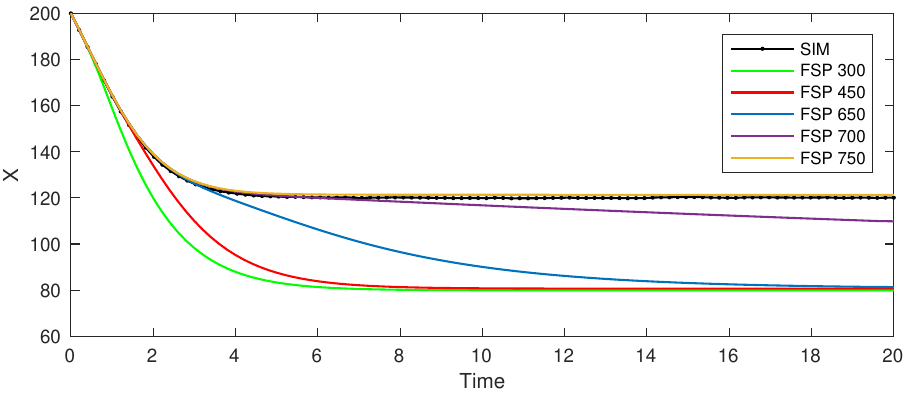}
\caption{{\bf Analysis by FSP of the Schl\"ogl system.} With parameter settings as in Fig.~\ref{fig:fse.example}, for state spaces of equal size as in Fig.~\ref{fig:fse.example}F (450 and 650), FSE estimates the average population of $X$ more precisely than finite state projection (stochastic simulation performed with \np{100000} runs). In particular, while FSE  requires a state space with 650 states to accurately match stochastic simulations, FSP needs 750 states.}\label{fig:schloegl.fsp}
\end{figure}
Defining incoming and outgoing transitions with respect to the buffer species maintained in the expanded network represents a crucial difference with FSP, where transitions toward unobserved state are collapsed into a sink state that absorbs the probability mass. Experimentally, this results in increased accuracy of mean estimates by FSE when tracking the same subset of the state space in both methods. For example,  Figure~\ref{fig:schloegl.fsp} shows this effect on the Schl\"ogl model; a similar behavior could be found in the other case studies (not reported here). More recently, variants of FSP have been proposed to cope with the decay of the probability mass, e.g., to obtain solutions that are  able to estimate stationary distributions and passage times\cite{doi:10.1137/15M1034180,doi:10.1063/1.5006484,doi:10.1063/5.0013457}. By means of a comparison against the most recent improvement by\cite{doi:10.1063/5.0013457}, in the enclosed Appendix 
(Section~\ref{sec:srm}) we show that FSE may provide accurate estimates using systems of fewer equations.

\begin{figure}
\centering
\begin{subfigure}[b]{0.495\linewidth}
\centering
\includegraphics[scale=0.40]{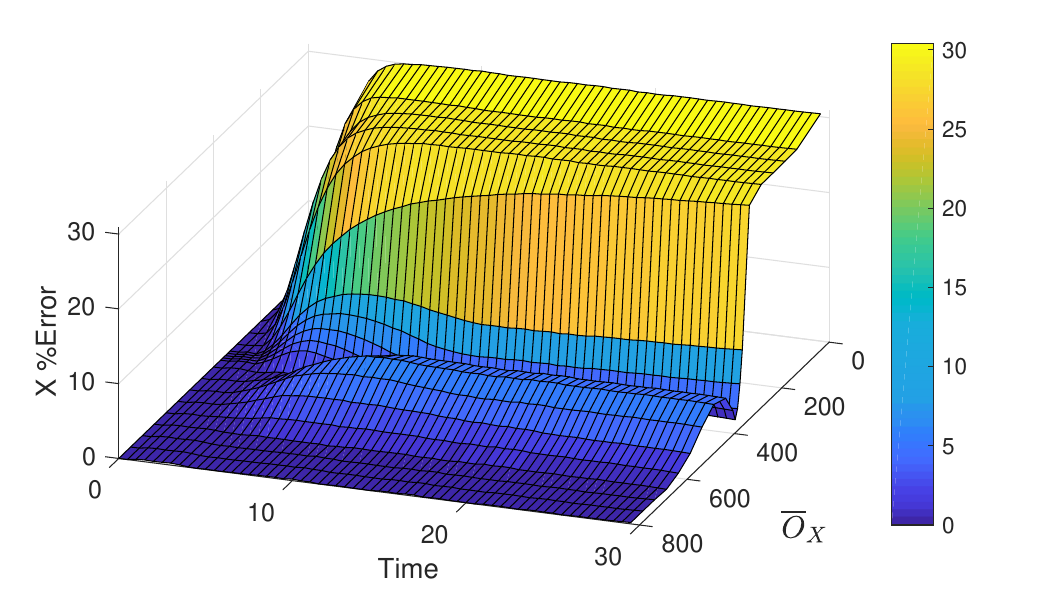}
\caption{{\bf Schl\"ogl system} (\np{100000} runs).}\label{fig:sen:schlogl}
\end{subfigure}
\hfill
\begin{subfigure}[b]{0.495\linewidth}
\centering
\includegraphics[scale=0.40]{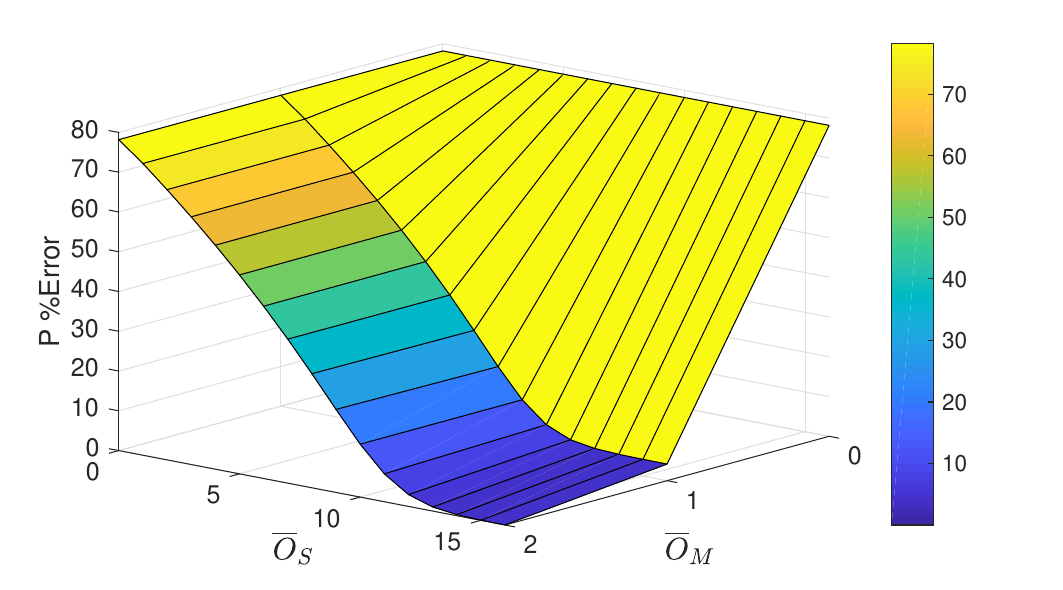}
\caption{{\bf Toggle switch} (\np{500000} runs, $t = 400$, $\DecompMax_P = 0$)}
\end{subfigure}
\begin{subfigure}[b]{0.98\linewidth}
\!\!\!\!\!\!\!\!\!\!\!\includegraphics[scale=1.25]{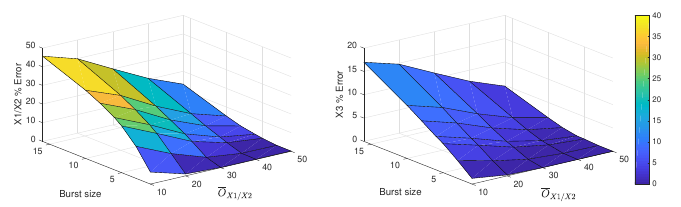}
\caption{{\bf Heterodimerization model} (\np{300000} runs, $t = 0.15$).}\label{fig:sen:hetero}
\end{subfigure}
\begin{subfigure}[b]{0.98\linewidth}
\!\!\!\!\!\!\!\!\!\!\!\includegraphics[scale=1.25]{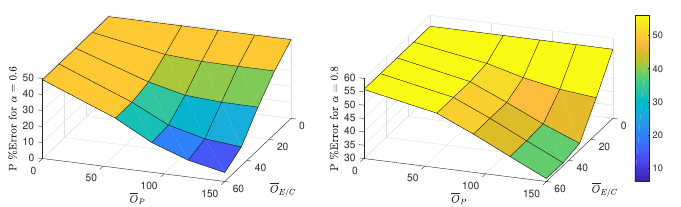}
\caption{{\bf Protein degradation model} (\np{200000} runs, $t = 2$).}
\end{subfigure}
\begin{subfigure}[b]{0.98\linewidth}
\!\!\!\!\!\!\!\!\!\!\!\includegraphics[scale=1.25]{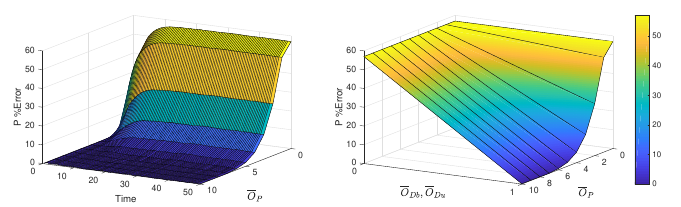}
\caption{ {\bf Feedback switch} (\np{1000000} runs); left: $\DecompMax_{D_b} =\DecompMax_{D_u}=1$; right: $t = 50$.}
\end{subfigure}
\caption{{\bf Sensitivity analysis}. Approximation error is measured as percentage error between the FSE estimate and the average by stochastic simulation. The reported time points are representative of steady state.}\label{fig:sen}
%
%
%
%
%
\end{figure}

The solution by the aforementioned state-space truncation methods is a lower bound on the true probability distribution, and increasing the set of observed states tightens that bound. Instead, although FSE ensures that the expansion coincides with the master equation when the whole state space is tracked, it does not give theoretical guarantees on the degree of accuracy, nor does it guarantee monotonically increasing accuracy with larger observation bounds. Indeed, experimentally we confirmed that monotonicity of the error is model dependent. For instance, the relative percentage error between the mean population predicted by FSE and the estimated mean by stochastic simulation is not monotonic in the Schl\"ogl system (Fig.~\ref{fig:sen}a),   
 while it is monotonic for the other case studies (Fig.~\ref{fig:sen}b-e).

\section{Conclusion}
We have presented finite state expansion (FSE) as a novel analytical method that offers a trade off between the exactness of the solution of the master equation and the approximation errors introduced by the deterministic rate equation (DRE) for stochastic reaction networks. FSE maintains a user-defined subset of the discrete state space and couples this with whole-population continuous dynamics to account for the behavior of states that are not explicitly tracked. By an algorithmic translation of a  network into an expanded one with auxiliary species and modified reactions, FSE leads to equations that can be interpreted as a refinement of the original DRE. A theoretical result of asymptotic correctness increases the confidence as to the effectiveness of the method, since it shows that the DRE of the expanded network corresponds to the original master equation. 

The performance of FSE in correcting the original DRE when tracking a strict subset of the discrete state space has been shown numerically in models that turn out to be challenging for related state-of-the-art techniques. The effective mesoscopic rate equation relies on perturbation arguments around the linear-noise approximation, hence it inherently assumes a limiting regime, unlike FSE. Experimentally, we found that this resulted in less accurate mean estimates than FSE. With respect to analytical approximations of the master equation based on moment closure, the case studies proved difficult since the analyses returned unphysical results or exhibited numerical issues, as also reported in the literature. 
Unlike state-space truncation methods based on finite state projection, FSE does not provide error bounds with respect to the original master equation. Yet, we found excellent accuracy when the observed state space is large enough.  
Overall, these findings make FSE a useful tool to study chemical reaction networks for which exact stochastic analysis through the master equation is not accessible. Currently, the FSE translation provides a refinement of mean estimates. It is an interesting line of future work to extend the method to compute improved approximations of higher-order moments. 

Despite the encouraging results herein reported, the applicability of FSE may not always be feasible. Since it is based on an enumeration of the discrete state space, FSE may too suffer from combinatorial complexity, such that the number of equations can grow rapidly large as a function of the observation bound. For larger networks in particular, this may require relatively small tracked state spaces to keep reasonable computational costs. On the other hand, if significant probability mass falls outside the tracked state space then the performance of FSE may not be adequate, as the Schl\"ogl system shows when small enough bounds are used. 

There are a number of methods that are worth investigating in the future in order to tackle these challenges.  Model-reduction techniques could help relieve the computational cost of the analysis of the DRE by providing a lower-order approximation that preserves the dynamics of interest~\cite{pnas17,10.1371/journal.pcbi.1005571}. It would be particularly beneficial to study either bounds or monotonicity properties on the FSE approximation error in order to develop adaptive strategies to find optimal values for the observation bounds. In principle, there might be other expansions than the one presented here, which give rise to different correction behavior of the DRE while still preserving the stochastic dynamics.  A further line of improvement might consist in devising variants of FSE where the tracked state space can be arbitrarily fixed, instead of being dependent on an upper bound for the population counts. This might allow the fine-tuning of the choice of the discrete region where the probability mass is mostly concentrated. For such expansions, smaller observation bounds (hence lower computational cost) may suffice to obtain the same degree of accuracy as in this paper, thus potentially extending the practical applicability of FSE to more complex models.\vskip6pt

\enlargethispage{20pt}

\ethics{This article does not present research with ethical considerations.}

\dataccess{Software implementation and all material to replicate the experiments herein reported are accessible from \href{https://zenodo.org/record/4739329}{https://zenodo.org/record/4739329}.}

\aucontribute{TW participated in the design of the study, carried out research, and drafted the manuscript; LB participated in the design of the study, helped draft and critically revise the manuscript; AV helped carry out the research and critically revised the manuscript; MT conceived the study, designed the study, coordinated the study and helped draft the manuscript. All authors gave final approval for publication and agree to be held accountable for the work performed therein.}

\competing{We declare we have no competing interests.}

\funding{This work has been partially supported by the Italian Ministry for Research (PRIN project ``SEDUCE'', no. 2017TWRCN).}



\bibliographystyle{RS} 
\bibliography{ms}
%
%
%
%
%

\appendix

\section{Proofs of statements}\label{sec:proofs}
This appendix provides accompanying material for the main text. This section shows the proofs of Theorems~\ref{thm:stochEqui} and~\ref{thm:dre} in the main text. Section~\ref{sec:srm} presents the comparison of finite state expansion (FSE) against the method of slack reactants~\cite{doi:10.1063/5.0013457}. 
\paragraph*{Notation.} We define the following operations  for any two $\sigma, \mu \in \mathbb{R}^\SpecSet$.
\begin{itemize}
	\item Minimum: $\sigma \land \mu$ is  such that  $\spop{S}{(\sigma \land \mu)} = \min(\spop{S}{\sigma}, \spop{S}{\mu})$ for all $S \in \SpecSet$.
	\item Saturated subtraction: $\sigma \ominus \mu$ is such that $\spop{S}{(\sigma\ominus \mu)} = \max(0, \spop{S}{\sigma} -\spop{S}{\mu})$ for all $S \in \SpecSet$. 
	\item Projection: Given $\mathscr{P} \subseteq \SpecSet$, $\sigma_{|\mathscr{P}} \in \mathbb{R}^\mathscr{P}$ is such that $({\sigma_{|\mathscr{P}}})_P = \sigma_P$ for all $P \in \mathscr{P}$. 
	\item Mapping: Given $\mathscr{P} \subseteq \SpecSet$,  
	and a function $m: \SpecSet \rightarrow \mathscr{P}$, $\sigma^m \in \mathbb{R}^\mathscr{P}$ is such that $\spop{P}{\sigma}^m = \sum_{m(S)=P} \spop{S}{\sigma}$, for all $P \in \mathscr{P}$.
\end{itemize}
With these, the quantities in Eq.~\ref{eq:eta} in the main text can be rewritten as follows: 
\begin{align*}
	\eta &= \rho \ominus o,  \label{eq:eta} &    
	\psi & =   \left((o \ominus \rho) + \pi\right) \ominus \DecompMax,  & 
	o' & = \DecompMax  \land \left((o \ominus \rho) + \pi\right).  
\end{align*}

\subsection{Proof of Theorem 3.1}
Let $P$ and $\hat{P}$ denote the solutions of the master equation in the original and expanded network, respectively. Then  
\[
\sum_{o + \xi = \sigma} \spop{\llbracket o\rrbracket  + \xi}{\hat{P}}(0) = \spop{\sigma}{P}(0)
\implies  \sum_{o + \xi = \sigma} \spop{\llbracket o\rrbracket  + \xi}{\hat{P}}(t) = \spop{\sigma}{P}(t),
\]
for all $t$.
\begin{proof}
	We prove the following equivalence for the derivatives of the solutions of the respective master equations
	\[  \sum_{o + \xi = \sigma} \frac{d\spop{\llbracket o\rrbracket  + \xi}{\hat{P}}}{dt} =  \frac{d\spop{\sigma}{{P}}}{dt} \qquad \text{for all~} \state \in \IntegerSpecies,
	\]
	from which the statement holds under the assumption of consistent initial conditions. 
	
	\[
	\sum_{o + \xi = \sigma} \frac{d\spop{\llbracket o\rrbracket  + \xi}{\hat{P}}}{dt} = 
	\]
	\begin{align}
		& = \sum_{o + \xi = \sigma} \sum_{(\llbracket \epsilon\rrbracket  + \eta) \xrightarrow{\ratef_\epsilon} (\llbracket \oi\rrbracket  + \psi) \in \ReacSet_{\DecompMax}} \!\!\!\!\!\!\!\!\!\!\!\!\!\!\!\!\!\!\!\!\!\!\!
		\Big(\ratef_\epsilon(\llbracket o\rrbracket  + \xi + \llbracket \epsilon\rrbracket + \eta \! - \llbracket \oi\rrbracket - \! \psi)   \hat{P}_{\llbracket o\rrbracket  + \xi + \llbracket \epsilon\rrbracket + \eta  - \llbracket \oi\rrbracket - \psi}
		- 
		\ratef_\epsilon(\llbracket o\rrbracket  + \xi) \!\cdot \!\hat{P}_{\llbracket o\rrbracket  + \xi}\Big) 
		\notag  \\
		& = \sum_{o + \xi = \sigma} \left( \sum_{(\llbracket \epsilon\rrbracket  + \eta) \xrightarrow{\ratef_\epsilon} (\llbracket o\rrbracket  + \psi) \in \ReacSet_{\DecompMax}} \!\!\!\!\!\!\!\!\!\!\!\!\!\!\!\!\!\!\!\!
		\ratef_\epsilon(\llbracket \epsilon
		\rrbracket  + \xi  + \eta  - \psi) \cdot \hat{P}_{\llbracket \epsilon\rrbracket  + \xi  + \eta  - \psi}
		- \!\!\!\!\!\!\!\!\!\!\!\!\!\!\!\!\!\!\!\!
		\sum_{(\llbracket o\rrbracket  + \eta) \xrightarrow{\ratef_o} (\llbracket \oi\rrbracket  + \psi) \in \ReacSet_{\DecompMax}} \!\!\!\!\!\!\!\!\!\!\!\!\!\!\!\!\!\!\!\!
		\ratef_o(\llbracket o\rrbracket  + \xi) \cdot \hat{P}_{\llbracket o\rrbracket  + \xi} \right) 
		\notag \\
		& =
		\sum_{\stackrel{(\llbracket \epsilon\rrbracket  + \eta) \xrightarrow{\ratef_\epsilon} (\llbracket o\rrbracket  + \psi) \in \ReacSet_{\DecompMax}} {o + \xi = \sigma}} \!\!\!\!\!\!\!\!\!\!\!\!\!\!\!\!\!
		\ratef_\epsilon(\llbracket \epsilon\rrbracket  + \xi  + \eta  - \psi) \cdot \hat{P}_{\llbracket \epsilon\rrbracket  + \xi  + \eta  - \psi}
		\ - \!\!\!\!\!\!\!\!\!\!\!\!\!\!\!\!\!\!
		\sum_{\stackrel{(\llbracket o\rrbracket  + \eta) \xrightarrow{\ratef_o} (\llbracket \oi\rrbracket  + \psi) \in \ReacSet_{\DecompMax}}{o + \xi = \sigma}} \!\!\!\!\!\!\!\!\!\!\!\!\!\!\!\!\!\!
		\ratef_o(\llbracket o\rrbracket  + \xi) \cdot \hat{P}_{\llbracket o\rrbracket  + \xi} 
		\notag\\
		& = \sum_{\stackrel{(\llbracket \epsilon\rrbracket  + \eta) \xrightarrow{\ratef_\epsilon} (\llbracket o\rrbracket  + \psi) \in \ReacSet_{\DecompMax}} {\epsilon + \xi + \eta - \psi = \sigma - (o+\psi) + (\epsilon+\eta)}} \!\!\!\!\!\!\!\!\!\!\!\!\!\!\!\!\!\!\!\!\!\!\!
		\ratef(\epsilon + \xi  + \eta  - \psi)\cdot 
		\underbrace{\spop{\llbracket \epsilon\rrbracket}{(\llbracket \epsilon\rrbracket  + \xi  + \eta  - \psi)}}_{=1} 
		\cdot \hat{P}_{\llbracket \epsilon\rrbracket  + \xi  + \eta  - \psi}
		\mathop{+} 
		\notag \\
		& \qquad \qquad \qquad \qquad \mathop{-} 
		\sum_{\stackrel{(\llbracket o\rrbracket  + \eta) \xrightarrow{\ratef_o} (\llbracket \oi\rrbracket  + \psi) \in \ReacSet_{\DecompMax}}{o + \xi = \sigma}} \!\!\!\!\!\!\!\!\!\!\!\!\!\!\!\!\!\!
		\ratef(o  + \xi) \cdot 
		\underbrace{\spop{\llbracket o\rrbracket}{(\llbracket o\rrbracket  + \xi)}}_{=1}\cdot \hat{P}_{\llbracket o\rrbracket  + \xi} 
		\notag\\
		& =  \sum_{\stackrel{\rho \xrightarrow{\ratef} \pi \in \ReacSet} {\epsilon + \xi + \eta - \psi = \sigma - \pi + \rho}} \!\!\!\!\!\!\!\!\!\!\!\!\! \ratef(\epsilon  + \xi  + \eta  - \psi) \cdot \hat{P}_{\llbracket \epsilon\rrbracket  + \xi  + \eta  - \psi}
		\ - \!\!
		\sum_{\stackrel{\rho \xrightarrow{\ratef} \pi \in \ReacSet}{o + \xi = \sigma}} \!\!\!\!
		\ratef(\sigma) \cdot \hat{P}_{\llbracket o\rrbracket  + \xi} 
		\notag\\ 
		& =  \sum_{\rho \xrightarrow{\ratef} \pi \in \ReacSet} \!\!
		\ratef(\sigma - \pi + \rho) \cdot P_{\sigma - \pi + \rho}
		\ - 
		\sum_{\rho \xrightarrow{\ratef} \pi \in \ReacSet}
		\ratef(\sigma) \cdot P_{\sigma} 
		\notag\\
		& = 
		\sum_{\rho \xrightarrow{\ratef} \pi \in \ReacSet} \!\! 
		\Big(\ratef(\sigma - \pi + \rho) \cdot P_{\sigma - \pi + \rho}
		\ - 
		\ratef(\sigma) \cdot P_{\sigma} \Big)
		\notag\\
		& =
		\quad \frac{d\spop{\sigma}{P}}{dt} . \notag
	\end{align}
\end{proof} 

Now we consider the other limit case, namely when the auxiliary set of species  contains all discrete states, corresponding to a fully expanded reaction network. In this case, the DRE of the expanded network corresponds to the master equation of the original network, hence no approximation occurs. 

\subsection{Proof of Theorem 3.2}

In order to prove Theorem~\ref{thm:dre} in the main text,  we prove two preliminary results stated as lemmata. 


\subsubsection*{Lemma A.1}
\begin{lemma}\label{thm:welldef}
	The expansion of a well-defined reaction network is well-defined.
\end{lemma}
\begin{proof}
	Let $\rho \xrightarrow{\ratef} \pi$ be a reaction of a well-defined network and $\llbracket o\rrbracket  + \eta \xrightarrow{\ratef_o} \llbracket \oi\rrbracket  + \psi$ its expansion according to the main text. Let us take  $z\in\ \mathbb{N}^{\SpecSet_{\DecompMax}}$ such that $(\llbracket o\rrbracket  + \eta)\nsubseteq z$ and separately consider the two cases for which this holds. If $\llbracket o\rrbracket\notin z$, then propensity function $\ratef_o$ in the expanded reaction is equal to zero by definition, keeping with the requirement for the reaction being well-defined.
	If $\eta\nsubseteq z$, then we have that $(\rho \ominus o) \nsubseteq z_{|\SpecSet}$ as $\rho$ and $o$ are both members of $\IntegerSpecies$. This implies that $\rho\nsubseteq(o+z_{|\SpecSet})$, and since, the reaction is well-defined we have that $\ratef(o+ z_{|\SpecSet})=0$, from which $\ratef_o( z) = 0$.
\end{proof}

\subsubsection*{Lemma A.2}
The following lemma proves that the expansion preserves the overall population jumps. That is, for each original reaction, every expanded reaction is such that each species is subject to the same change of its abundance level.  

\begin{lemma}\label{thm:totalChange}
	Let $\rho \xrightarrow{\ratef} \pi$ be a reaction and $\llbracket o\rrbracket  + \eta \xrightarrow{\ratef_o} \llbracket \oi\rrbracket  + \psi$ its expansion according to 
	The main text. 
	Then it holds that:
	\begin{enumerate}
		\item $(o + \eta) \ominus (\oi  +  \psi) = \rho \ominus \pi$;
		\item $(\oi + \psi) \ominus (o + \eta) = \pi \ominus \rho$;
		\item $\sigma \ominus (o + \eta) + (\oi  +  \psi) = \sigma \ominus \rho + \pi$, for all $\sigma \in \mathbb{N}^\SpecSet$ such that $(o + \eta) \subseteq \sigma$.
	\end{enumerate}
\end{lemma}


%
\begin{proof}
	For case (1):
	\begin{align*}
		(o + \eta) \ominus (\oi  +  \psi) 
		= & (o + (\rho \ominus o)) \ominus ((\DecompMax \land (o \ominus \rho + \pi) \\
		& \qquad + ((o \ominus \rho + \pi) \ominus \DecompMax) ) \\
		= & (o + (\rho \ominus o)) \ominus (o \ominus \rho + \pi) \\
		= & (\rho + (o \ominus \rho)) \ominus ((o \ominus \rho) + \pi) \\
		= & \  \rho \ominus \pi .
	\end{align*}
	
	\noindent For case (2): 
	\begin{align*}
		(o'  +  \psi) \ominus (o + \eta) 
		= &  ((\DecompMax \land (o \ominus \rho + \pi) + ((o \ominus \rho + \pi) \ominus \DecompMax) ) \\
		& \qquad \ominus (o + (\rho \ominus o)) \\
		= &   (o \ominus \rho + \pi) \ominus (o + (\rho \ominus o)) \\
		= &  ((o \ominus \rho) + \pi) \ominus (\rho + (o \ominus \rho)) \\
		= & \  \pi \ominus \rho .
	\end{align*}
	
	\noindent For case (3): 
	\begin{align}
		\sigma \ominus \rho + \pi 
		&=\sigma \ominus (\rho \ominus \pi) + (\pi \ominus \rho) \notag \\
		&= \sigma \ominus ((o + \eta) \ominus (\oi  +  \psi)) + (\oi + \psi) \ominus (o + \eta) \notag \\
		&= \sigma \ominus ((o + \eta) \ominus( (o + \eta) \land (\oi  +  \psi))) \notag \\
		& \qquad + (\oi + \psi) \ominus ((o + \eta)\land (\oi  +  \psi)) \label{eq:a} \\
		&= \sigma \ominus (o + \eta) + ((o + \eta) \land (\oi  +  \psi))  \notag \\
		& \qquad + (\oi + \psi) \ominus ((o + \eta)\land (\oi  +  \psi)) \label{eq:b} \\
		&= \sigma \ominus (o + \eta) + (\oi + \psi) + ((o + \eta) \land (\oi  +  \psi))\notag \\
		& \qquad \ominus ((o + \eta)\land (\oi  +  \psi))\notag \\
		&= \sigma \ominus (o + \eta) + (\oi + \psi), \notag 
	\end{align}
	where Eq.~(\ref{eq:b}) follows from Eq.~(\ref{eq:a}) because of the relations:
	\[o + \eta \leq \sigma \ \text{~and~} \  (o + \eta) \geq (o + \eta) \land (\oi  +  \psi) \leq (\oi  +  \psi). \]
\end{proof}

\subsubsection*{Proof of Theorem 3.2}
Consider a well-defined reaction network $(\SpecSet,\ReacSet)$ and let $(\SpecSet_{\DecompMax}, \ReacSet_{\DecompMax})$ 
be its expansion 
where
$$\SpecSet_{\DecompMax} = \SpecSet \cup \left\{ \llbracket o \rrbracket \mid o \in \IntegerSpecies \right\}.$$
Let $X(t)$ be the \dre\ solution of the expanded network and $P(t)$ the solution of the master equation of the original network at time $t$.  
Then
\begin{enumerate}
	\item if $X_S(0)= 0$ then $X_S(t)= 0$ for all $t$ and $S \in \SpecSet$;
	\item if $X_{\tracked{o}}(0) = P_o(0)$ then $X_{\llbracket o \rrbracket}(t) = P_o(t)$, for all $t$ and $o \in \IntegerSpecies$.
\end{enumerate}
\begin{proof}
	Case i). This statement holds if, whenever $X_S(t) = 0$, then $\frac{d\spop{S}{X}(t)}{dt}=0$ for all $S \in \SpecSet$. The \dre\ for the expanded reaction network can be written as follows:
	\begin{equation}\label{eq:dre.decomp} 
		\frac{d\spop{S}{X}}{dt} = \sum_{\rho \xrightarrow{\ratef_o} \pi \in \ReacSet_{\DecompMax}} (\spop{S}{\pi} - \spop{S}{\rho}) \cdot \ratef_o(X), \qquad \text{for all~} S \in \SpecSet.
	\end{equation}
	Since $o \in \IntegerSpecies$, every expanded reaction ${\llbracket o\rrbracket} + \eta \xrightarrow{~f_o~} {\llbracket o'\rrbracket} + \psi$ will be such that $\psi_S = 0$ for each $S \in \SpecSet$, hence $\pi_S = 0$ in Eq.~(\ref{eq:dre.decomp}). 
	Let us now assume toward a contradiction that  $\frac{d\spop{S}{X}(t)}{dt} \neq 0$ for $X_S(t) = 0$. This must hold only if both $\rho_S\neq 0$ and $\ratef_o(X(t))\neq 0$ for a  reaction $\rho \xrightarrow{\ratef_o} \pi \in \ReacSet_{\DecompMax}$ expanded from $\rho' \xrightarrow{\ratef} \pi' \in \ReacSet$. For a given auxiliary species $\llbracket o\rrbracket$, the propensity function is in the form $f_o(X(t)) = \spop{\llbracket o\rrbracket}{X}(t) \cdot \ratef(o + X_{|\SpecSet}(t))$. Since  $X_S(t) =0$ for each $S \in \SpecSet$ this reduces to $f_o(X(t)) = \spop{\llbracket o\rrbracket}{X}(t) \cdot \ratef(o)$.
	As the reaction network is well-defined, 
	$f(o)>0$ implies that $\rho'\leq o$.
	In this case, from Eq~\ref{eq:eta} in the main text it follows that $\rho$ must be in the form $\rho = \llbracket o\rrbracket+\emptyset$, that is, $\rho_S=0$ for all $S\in\SpecSet$, closing this case by contradiction.
	
	\noindent  Case ii). For each $\llbracket o\rrbracket$, the \dre\ can be written as: 
	\begin{align*} 
		\frac{d\spop{\llbracket o\rrbracket}{X}}{dt} 
		=& \sum_{(\llbracket \epsilon\rrbracket  + \rho) \xrightarrow{\ratef_\epsilon} (\llbracket o\rrbracket  + \xi) \in \ReacSet_{\DecompMax}} \ratef_\epsilon(X)  \\
		& \qquad - \sum_{(\llbracket o\rrbracket  + \xi) \xrightarrow{\ratef_o} (\llbracket \oi\rrbracket  + \pi) \in \ReacSet_{\DecompMax}}  \ratef_o(X) \\
		=& \sum_{(\llbracket \epsilon\rrbracket  + \rho) \xrightarrow{\ratef_\epsilon} (\llbracket o\rrbracket  + \xi) \in \ReacSet_{\DecompMax}} \spop{\llbracket \epsilon\rrbracket}{X} \cdot \ratef(\epsilon + X_{|\SpecSet})  \\
		& \qquad 
		- \sum_{(\llbracket o\rrbracket  + \xi) \xrightarrow{\ratef_o} (\llbracket \oi\rrbracket  + \pi) \in \ReacSet_{\DecompMax}}  \spop{\llbracket o\rrbracket}{X} \cdot \ratef(o + X_{|\SpecSet}).
	\end{align*}
	Since $X_S = 0$ and the expanded reaction network is well-defined by Lemma~\ref{thm:welldef}, this simplifies to:
	\begin{align*} 
		\frac{d\spop{\llbracket o\rrbracket}{X}}{dt} 
		=& \sum_{(\llbracket \epsilon\rrbracket  + \emptyset) \xrightarrow{\ratef_\epsilon} (\llbracket o\rrbracket  + \emptyset) \in \ReacSet_{\DecompMax}} \spop{\llbracket \epsilon\rrbracket}{X} \cdot \ratef(\epsilon + \emptyset)  \\ 
		& \qquad - \sum_{(\llbracket o\rrbracket  + \emptyset) \xrightarrow{\ratef_o} (\llbracket \oi\rrbracket  + \emptyset) \in \ReacSet_{\DecompMax}}  \spop{\llbracket o\rrbracket}{X} \cdot \ratef(o + \emptyset) \\
		=& \sum_{(\llbracket \epsilon\rrbracket  + \emptyset) \xrightarrow{\ratef_\epsilon} (\llbracket o\rrbracket  + \emptyset) \in \ReacSet_{\DecompMax}} \spop{\llbracket \epsilon\rrbracket}{X} \cdot \ratef(\epsilon)  \\ 
		& \qquad - \sum_{(\llbracket o\rrbracket  + \emptyset) \xrightarrow{\ratef_o} (\llbracket \oi\rrbracket  + \emptyset) \in \ReacSet_{\DecompMax}}  \spop{\llbracket o\rrbracket}{X} \cdot \ratef(o).
	\end{align*}
	The summations in the above equation can be written in terms of reactions of the original network as follows:
	\begin{align*} 
		\frac{d\spop{\llbracket o\rrbracket}{X}}{dt} 
		=& \sum_{(\llbracket \epsilon\rrbracket  + \emptyset) \xrightarrow{\ratef_\epsilon} (\llbracket o\rrbracket  + \emptyset) \in \ReacSet_{\DecompMax}} \spop{\llbracket \epsilon\rrbracket}{X} \cdot \ratef(\epsilon)  \\
		& \qquad  - \sum_{(\llbracket o\rrbracket  + \emptyset) \xrightarrow{\ratef_o} (\llbracket \oi\rrbracket  + \emptyset) \in \ReacSet_{\DecompMax}}  \spop{\llbracket o\rrbracket}{X} \cdot \ratef(o) \\
		=& \sum_{\stackrel{\rho \xrightarrow{\ratef} \pi \in \ReacSet}{ o = \epsilon \ominus \rho + \pi }} \spop{\llbracket \epsilon\rrbracket}{X} \cdot \ratef(\epsilon)  \ \ 
		- \sum_{\stackrel{\rho \xrightarrow{\ratef} \pi \in \ReacSet}{ \oi = o \ominus \rho + \pi}}  \spop{\llbracket o\rrbracket}{X} \cdot \ratef(o)
		\\
		=& \sum_{\stackrel{\rho \xrightarrow{\ratef} \pi \in \ReacSet}{ \epsilon = o + \rho - \pi }} \spop{\llbracket \epsilon\rrbracket}{X} \cdot \ratef(\epsilon)  \ \ 
		- \sum_{\rho \xrightarrow{\ratef} \pi \in \ReacSet}  \spop{\llbracket o\rrbracket}{X} \cdot \ratef(o)
		\\
		=& \sum_{\rho \xrightarrow{\ratef} \pi \in \ReacSet} \!\! 
		\Big(
		\ratef(o - \pi + \rho) \cdot X_{\llbracket o - \pi + \rho\rrbracket}
		\ - 
		\ratef(o) \cdot X_{\llbracket o\rrbracket} \Big),
	\end{align*}
	from which the claim follows by noting that substituting variable name $X_{\llbracket \sigma\rrbracket}$ with $P_\sigma$ , for all $\sigma\in\IntegerSpecies$, precisely gives the original master equation.
\end{proof}

\section{Comparison against the linear mapping approximation}\label{sec:lma}
\begin{figure}
	\centering
	\includegraphics[scale=0.63]{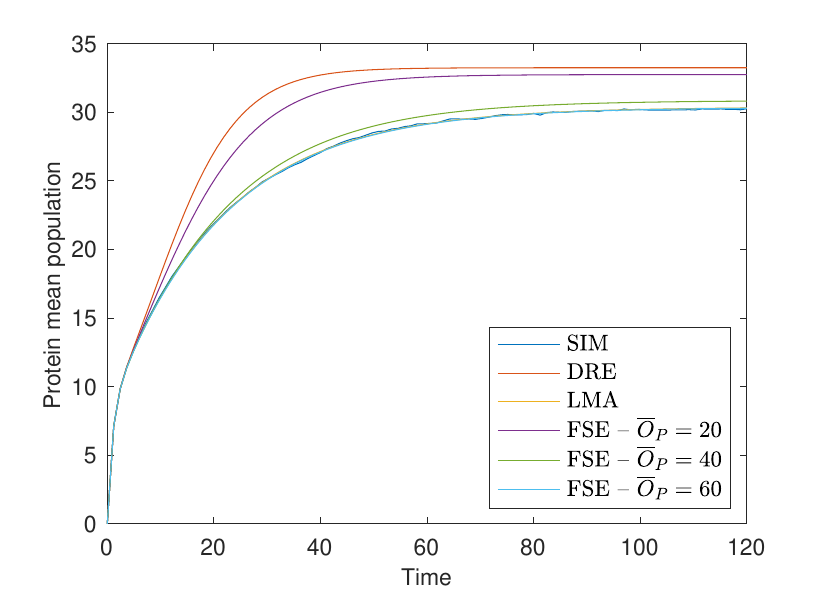}
	\caption{Comparison between FSE and linear mapping approximation (LMA) using the example in~\cite[Fig. 1]{lma} with parameter settings given by `Point B' as therein specified. The model corresponds to the network from Eq.~\ref{eq:feedback_switch} with $k_b$ = 0. The plot shows the average population of the protein species $P$ (computed by stochastic simulation using 50k runs); the DRE approximation; the LMA approximation; and the approximation by FSE by setting $\overline{O}_{P} = 20, 40, 60$ and fixing $\overline{O}_{D_u} = \overline{O}_{D_b} = 1$. The trajectories for LMA and FSE with the largest observation bound $\overline{O}_{P} = 60$ provide an excellent agreement with stochastic simulation; however, FSE requires 65 equations, whereas LMA only defines 3 linear differential equations to provide a refined estimate of the mean protein populations.}
	\label{fig:lma}
\end{figure}
This section compares FSE against  linear mapping approximation~\cite{lma}, an analytical technique to compute probability distributions of species populatuions in a class of models of stochastic gene expression. In particular we consider the feedback switch model from~\cite[Fig. 1]{lma}, which turns out to correspond to the network from Eq.~\ref{eq:feedback_switch} with $k_b$ = 0. Figure~\ref{fig:lma} confirms that FSE can accurately approximate the mean population of the protein if the observation bound for the number of proteins is large enough, i.e., 60 in this case, corresponding to a system of (nonlinear) ODEs with 65 variables. On the other hand, LMA can achieve a similar accuracy, overlapping the simulated mean trajectory, in a more parsimonious fashion with a system of 3 linear ODEs.

\section{Comparison against the slack reactants method}\label{sec:srm}
This section compares FSE against recent improvements of the finite state projection method (discussed in the main text) which cope with the problem of leaking probability mass into the absorbing state used for the truncation of the original state space. In particular we consider a numerical comparison against the most recently published method, called the \emph{slack reactants} method (SRM)~\cite{doi:10.1063/5.0013457}, which has been shown to offer the best performance against the other state-space truncation techniques. 

The idea of SRM is to add ``buffer species'' that turn a reaction network with an infinite state space into one with a finite state space, but without absorbing states, while still preserving convergence properties as with the original FSP. Eliminating absorbing states allows for stochastic analysis over long time horizons, such as first passage times and estimations of stationary distributions. 

\begin{figure}[t]
	\centering
	\includegraphics[scale=0.63]{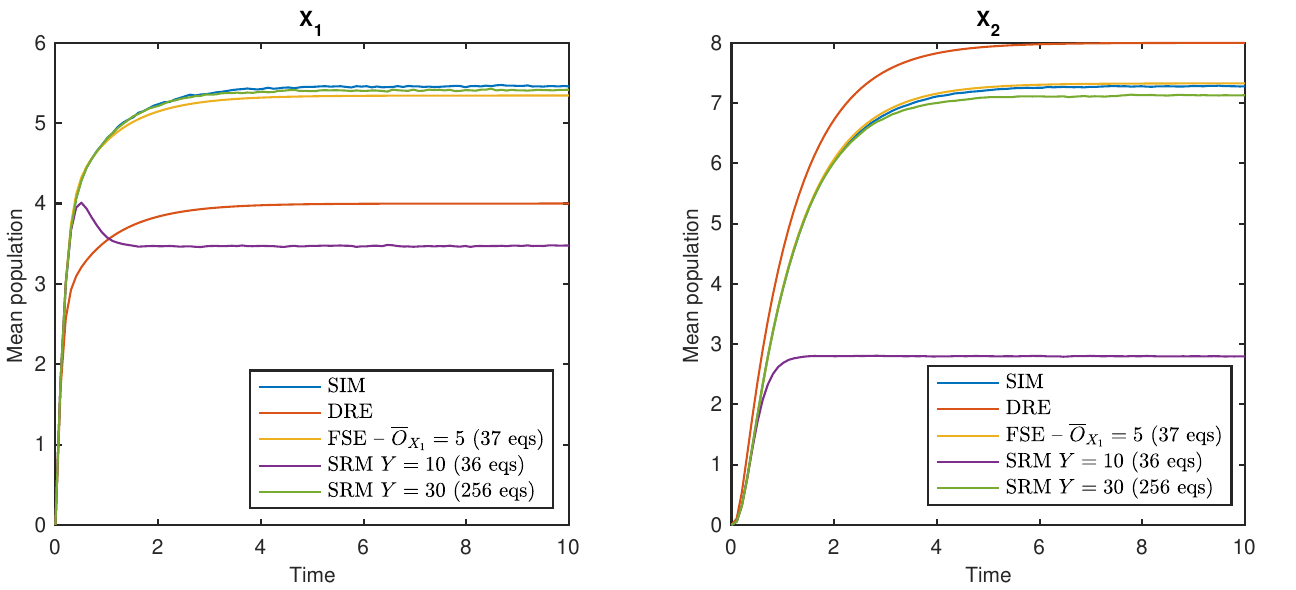}
	\caption{Comparison between FSE and SRM using the example in~\cite{doi:10.1063/5.0013457}. The plots show the average populations of species $X_1$ and $X_2$ (computed by stochastic simulation); their DRE approximation; the approximation by FSE by setting $\overline{O}_{X_1} = \overline{O}_{X_2} = 5$; and two estimates by SRM by setting the initial population of the buffer species $Y$ equal to 10 and 30, respectively (with the master equation solved by stochastic simulation). The SRM setting $Y=10$ gives a comparable number of equations than the FSE setting; however, the accuracy of the mean approximation is worse than DRE. On the other hand, the SRM setting $Y=30$ gives estimates of comparable accuracy to those by FSE, but it gives rise to 256 equations instead of 37 as in FSE.}\label{fig:srm}
\end{figure}
We compare FSE against SRM an example provided by the authors of SRM in~\cite[Fig. 3A]{doi:10.1063/5.0013457}. The chemical reaction network is reported here for convenience:
\begin{align*}
	X_1 & \xrightarrow{1} \emptyset & \emptyset & \xrightarrow{20} X_1 \\
	X_1 + X_1 & \xrightarrow{1} X_2 & X_2 & \xrightarrow{1} X_1 + X_1 \\
	X_2 & \xrightarrow{1} \emptyset
\end{align*}  
SRM modifies these reactions by adding a further species $Y$ that acts as a buffer for the copies of the original species $X_1$ and $X_2$ that enter or exit the network. The modified reaction network is as follows:
\begin{align*}
	X_1 & \xrightarrow{1} Y & Y & \xrightarrow{f} X_1 \\
	X_1 + X_1 & \xrightarrow{1} X_2 & X_2 & \xrightarrow{1} X_1 + X_1 \\
	X_2 & \xrightarrow{1} Y+Y 
\end{align*}  
where $f$ is a modified propensity function of the form $20\min(Y,1)$.

The stochastic behavior of this network increasingly corresponds to that of the original one with larger buffer sizes represented by the initial number of copies of $Y$; on the other hand, the larger such initial population the larger the number of states in the now finite state space of the Markov chain (hence the larger the number of equations in the master equation).  
Thus, a comparison between FSE and SRM can be done by measuring the difference between the estimations of the average populations given by the solution of the master equation of SRM and the DRE of FSE when both methods are set such that they give rise to systems of equations of similar size. Fig.~\ref{fig:srm} shows the results of this comparison, indicating that SRM needs ca. 7 times more equations to estimate the average population of the original species $X_1$ and $X_2$ in this model. 

\end{document}